\documentclass[seceq]{ptptex}

\usepackage{graphicx}

\def\tr{\mathop{\rm tr}\nolimits}
\def\mod{\mathop{\rm mod}\nolimits}
\def\Vol{\mathop{\rm Vol}}

\newcommand{\wt}{\tilde}

\preprintnumber[3cm]{
UT-08-32}

\title{
${\cal N}=4$ Chern-Simons theories and\\wrapped M-branes in their gravity duals%
}

\author{
Yosuke \textsc{Imamura}\thanks{E-mail: \tt imamura@hep-th.phys.s.u-tokyo.ac.jp}
 and Shuichi \textsc{Yokoyama}\thanks{E-mail: \tt yokoyama@hep-th.phys.s.u-tokyo.ac.jp}%
}

\inst{
Department of Physics, University of Tokyo,\\
Hongo 7-3-1, Bunkyo-ku, Tokyo 113-0033, Japan
}

\abst{
We investigate a class of ${\cal N}=4$ quiver Chern-Simons
theories and their gravity duals.
We define the group of fractional D3-brane charges in
a type IIB brane setup with taking account of
D3-brane creation due to Hanany-Witten effect,
and confirm that it agrees with the $3$-cycle homology of the dual geometry,
which describes the charges of fractional M2-branes,
M5-branes wrapped on $3$-cycles.
The relation between the fractional brane charge and the torsion of the
three-form potential field is partially established.
We also discuss the duality between baryonic operators
in the Chern-Simons theories
and M5-branes wrapped on $5$-cycles in
the dual geometries.
The degeneracy and the conformal dimension
of the operators
are reproduced on the gravity side.
We also comment on the relation between wrapped M2-branes and monopole operators.
The baryonic operators we consider are not gauge invariant.
We argue that
the gauge invariance cannot be imposed
on all the operators corresponding to wrapped M-branes
in $AdS_4/CFT_3$ correspondence.}

\begin{document}

\maketitle

\section{Introduction}
Recently, three-dimensional Chern-Simons matter systems
with various numbers of supersymmetries have attracted a great interest
as theories describing the low energy effective theories of M2-branes
in various backgrounds.
Bagger, Lambert, and Gustavsson
\cite{Bagger:2006sk,Bagger:2007jr,Bagger:2007vi,Gustavsson:2007vu,Gustavsson:2008dy} proposed an ${\cal N}=8$ Chern-Simons theory
as a model for multiple M2-branes.
This model (BLG model) 
is based on an interesting mathematical structure, so-called $3$-algebra.
The consistency condition (the fundamental identity)
associated with the $3$-algebra
is very restrictive, and there is only one example of superconformal
theory based on BLG model, which describes
two M2-branes in certain orbifolds \cite{Lambert:2008et,Distler:2008mk}.

After this proposal, many works about superconformal Chern-Simons
theories appeared.
In \citen{Gaiotto:2008sd}, a new class of Chern-Simons matter systems
which possess ${\cal N}=4$ supersymmetry is constructed.
It is generalized in \citen{Hosomichi:2008jd} by introducing new
matter multiplets (twisted hyper multiplets).

A theory describing multiple M2-branes in the eleven-dimensional flat spacetime
was first proposed by Aharony, Bergman, Jafferis,
and Maldacena \cite{Aharony:2008ug}.
Their model (ABJM model) is a $U(N)\times U(N)$ Charn-Simons theory
at level $(k,-k)$, and possesses ${\cal N}=6$ supersymmetry.
They show that the model describes multiple M2-branes in the orbifold ${\bf C}^4/{\bf Z}_k$.
If we take $k=1$, this space becomes ${\bf C}^4$,
and the supersymmetry is expected to be enhanced to ${\cal N}=8$ in some
non-trivial way.
See also \citen{Hosomichi:2008jb,Bagger:2008se,Schnabl:2008wj} for
${\cal N}\geq4$ Chern-Simons theories.

We can realize ABJM model by using a type IIB brane configuration
consisting of D3-branes, one NS5-brane and one $(k,1)$-fivebrane \cite{Aharony:2008ug}.
This is generalized by increasing the number of fivebranes
to quiver Chern-Simons theories with circular quiver diagrams \cite{Imamura:2008nn}.
If we introduce only two kinds of fivebranes with appropriate directions,
the Chern-Simons theory possesses ${\cal N}=4$ supersymmetry \cite{Imamura:2008dt}.
In this case, the corresponding geometry is a certain orbifold of ${\bf C}^4$ \cite{Imamura:2008nn}.
See also \citen{Benna:2008zy,Terashima:2008ba} for orbifolds of ABJM model.
If we introduce fivebranes with three or more different charges,
the number of supersymmetry is at most ${\cal N}=3$,
and we have in general curved hyper-K\"ahler geometries \cite{Jafferis:2008qz}.

All these theories are conformal
and have gravity duals \cite{Maldacena:1997re}.
The purpose of this paper is to investigate some aspects
about the gravity duals of ${\cal N}=4$ Chern-Simons theories.

One is about fractional branes.
Fractional branes in ABJM model are investigated in \citen{Aharony:2008gk}.
It is suggested that in the gravity dual the fractional branes are
realized as the torsion of the $3$-form potential in the orbifolds
${\bf S}^7/{\bf Z}_k$.
In this paper we generalize this result to more general orbifold
$M_{p,q,k}=({\bf S}^7/({\bf Z}_p\oplus{\bf Z}_q))/{\bf Z}_k$ associated with
${\cal N}=4$ Chern-Simons theories.
We find that the homology $H_3$ of this orbifold
is pure torsion again, and
confirm that it agrees with the group of
fractional brane charges in the type IIB brane configuration
obtained by taking account of Hanany-Witten effect.

We also discuss baryonic operators and monopole operators.
The gauge symmetry of the ${\cal N}=4$ Chern-Simons theory is
\begin{equation}
G
=(\prod_{I=1}^n
U(N_I)_I)/U(1)_d
=G_{SU}\times G_B,
\label{gg}
\end{equation}
where $U(1)_d$ is the diagonal $U(1)$ subgroup which does not couple to
any bi-fundamental fields,
and $G_{SU}$ and $G_B$ are defined by
\begin{equation}
G_{SU}=\prod_{I=1}^n SU(N_I)_I,\quad
G_B=(\prod_{I=1}^n U(1)_I)/U(1)_d.
\label{ggs}
\end{equation}
The abelian part $G_B=U(1)^{n-1}$
is called baryonic symmetry.
In the case of four-dimensional quiver gauge theories,
the baryonic symmetry is often treated as a global symmetry
because in the infra-red limit it decouples from the system.
In this paper, we treat this part of gauge group in
a similar way.
Namely, when we define baryonic operators later,
we require gauge invariance with respect only to the
$G_{SU}$ part of the gauge group.
Unlike the four-dimensional case,
$G_B$ does not decouple even in the infra-red limit,
and thus baryonic operators we discuss in this paper
are not gauge invariant operators.
In the case of ABJM model, the baryonic symmetry
$G_B=U(1)$ is spontaneously broken by the vacuum expectation values of
the dual photon field, and
we can define gauge invariant baryonic operators
by multiplying appropriate functions of the dual photon field\cite{Park:2008bk}.
This, however, does not work for $n\geq3$.

Despite the gauge variance of baryonic operators,
we identify them with M5-branes wrapped on fivecycles
in $M_{p,q,k}$,
and confirm that the conformal dimension and multiplicity
of the operators are reproduced
on the gravity side in the same way
as \citen{Berenstein:2002ke}, in which
the baryonic operators \cite{Gubser:1998fp}
in the Klebanov-Witten theory \cite{Klebanov:1998hh}
are investigated.
This may seem contrary to the usual AdS/CFT dictionary,
which relates only gauge invariant operators to counterparts on
the gravity side.
We discuss this point after mentioning the relation
between monopole operators and wrapped M2-branes.
In three-dimentional spacetime local operators in general carry magnetic 
charges. 
A special class of monopole operators constructed with the dual photon field
carry the magnetic charge of the diagonal $U(1)$ gauge group.
We thus call them diagonal monopole operators.
When $n\geq3$ there is more variety of monopole operators
in addition to diagonal ones.
We propose that such non-diagonal monopole operators
correspond to M2-branes wrapped on two-cycles in the internal space.

The rest of this paper is organized as follows.
In the next section we summarize field contents,
symmetries, and the moduli space of ${\cal N}=4$ Chern-Simons
theories.
In section \ref{hw.sec},
we determine the group of fractional branes for $k=1$
by using the type IIB setup,
and in Section \ref{frac.sec} we reproduce it as
the homology on the M-theory side.
We discuss the correspondence between baryonic operators
in Chern-Simons theories and M5-branes wrapped on
five-cycles for $k=1$ in Section \ref{baryon.sec}.
The analysis in Sections \ref{hw.sec}, \ref{frac.sec}, and \ref{baryon.sec}
are generalized to $k\geq2$ in Section \ref{kgeq2.sec}.
In Section \ref{qb.sec} we again discuss relation between
baryonic operators and wrapped M5-branes.
We show there that in the type IIB setup $N$ open strings representing constituent
bi-fundamental quarks can be continuously deformed into a D3-brane disk,
which is dual to a wrapped M5-brane.
In Section \ref{torsion.sec}, we discuss the relation
between fractional branes and torsion of the three-form potential
in the M-theory background.
In Section \ref{wrappedm2.sec}, we comment on the relation between
wrapped M2-branes and non-diagonal monopole operators,
and explain why we do not impose $G_B$ gauge invariance on baryonic operators.
We conclude the results in the last section.

\section{${\cal N}=4$ Chern-Simons theories}
Let us consider an ${\cal N}=4$ supersymmetric Chern-Simons theory
with gauge group (\ref{gg}).
This theory
includes the same number of
vector multiplets and bi-fundamental hypermultiplets,
and is represented by a circular quiver diagram.
\begin{figure}[htb]
\centerline{\includegraphics{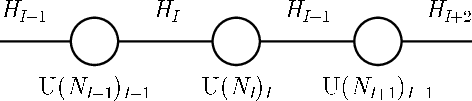}}
\caption{A part of a circular quiver diagram of an ${\cal N}=4$ supersymmetric Chern-Simons theory is shown.}
\label{quiver.eps}
\end{figure}
The size of gauge groups
$N_I$ may depend on vertices.

A hypermultiplet $H_I$ contains two complex scalar fields and
they belong to a doublet of $SU(2)$ R-symmetry.
The R-symmetry $Spin(4)=SU(2)^2$ of the ${\cal N}=4$ theory includes
two $SU(2)$ factors, and correspondingly there are two
kinds of hypermultiplets,
which are called untwisted and twisted hypermultiplets \cite{Hosomichi:2008jd}.
We denote two $SU(2)$ groups by $SU(2)_A$ and $SU(2)_B$,
and adopt the convention in which scalar components
of untwisted and twisted hypermultiplets are non-trivially transformed
by $SU(2)_A$ and $SU(2)_B$, respectively.
We denote these scalar fields by $h_I^\alpha$ and $h_I^{\dot\alpha}$.
(Undotted and dotted indices are ones for $SU(2)_A$ and $SU(2)_B$, respectively.)
Fermions are transformed in the opposite way from the scalar fields.
The theory also possesses two $U(1)$ global symmetries as is shown in
Table \ref{hyper.tbl}.
\begin{table}[htb]
\caption{The global symmetries of the ${\cal N}=4$ Chern-Simons theory are shown.
A certain combination of $U(1)_b$ and $U(1)_d$ acts on the hypermultiplets
in the same way as
a certain $U(1)$ subgroup of gauge symmetry.
This, however, does not act on the dual photon field,
and, by this reason, is different from the gauge symmetry.}
\label{hyper.tbl}
\begin{center}
\begin{tabular}{ccccc}
\hline
\hline
            &\multicolumn{2}{c}{untwisted hypermultiplets}&
            \multicolumn{2}{c}{twisted hypermultiplets}\\
           & $h_I^\alpha$ & $\psi_I^{\dot\alpha}$
           & $h_I^{\dot\alpha}$ & $\psi_I^\alpha$ \\
\hline
$SU(2)_A$ & ${\bf 2}$ & ${\bf 1}$ & ${\bf 1}$ & ${\bf 2}$ \\
$SU(2)_B$ & ${\bf 1}$ & ${\bf 2}$ & ${\bf 2}$ & ${\bf 1}$ \\
$U(1)_b$ & $+1$ & $+1$ & $-1$ & $-1$ \\
$U(1)_d$ & $+1$ & $+1$ & $+1$ & $+1$ \\
\hline
\end{tabular}
\end{center}
\end{table}

In the type IIB brane system, which consists of
D3-branes, NS5-branes, and $(k,1)$-fivebranes,
these hypermultiplets arises from the open strings stretched between
two adjacent intervals of D3-branes.
We use the same index $I$ for fivebranes as hypermultiplets.
The two kinds of hypermultiplets correspond to the
two different charges of fivebranes.
Let us define numbers $s_I$ associated with hypermultiplets
which are $0$ for untwisted hypermultiplets and $1$ for twisted hypermultiplets.
The RR charge of the fivebrane associated with the $I$-th hypermultiplet
is $ks_I$,
and the boundary interaction of D3-branes ending on fivebranes
induces the Chern-Simons terms \cite{Kitao:1998mf,Bergman:1999na}
\begin{equation}
S=\sum_I\frac{k_I}{4\pi}\int\tr\left(A_IdA_I+\frac{2}{3}A_I^3\right),
\end{equation}
where the level of the $U(N_I)_I$ gauge group
coupling to the hypermultiplets $H_I$ and $H_{I+1}$
is given by
\begin{equation}
k_I=k(s_{I+1}-s_I).
\end{equation}
In the following, we refer to the integer $k$ simply as the ``level'' of the theory.

The moduli space of this theory is analyzed in \citen{Imamura:2008nn}.
We obtain the background geometry of M2-branes
as the Higgs branch moduli space of
the theory with $N_I=1$.
When $k=1$, it is the product of two four-dimensional orbifolds.
\begin{equation}
{\cal M}_{p,q}
={\bf C}^2/{\bf Z}_p
\times{\bf C}^2/{\bf Z}_q,
\label{orbi}
\end{equation}
where $p$ and $q$ are the numbers of untwisted and twisted
hypermultiplets, respectively.
For later convenience, we introduce complex coordinates
$z_i$ ($i=1,2,3,4$) on which the orbifold group acts by
\begin{eqnarray}
(z_1,z_2,z_3,z_4)
\rightarrow (\omega_p^mz_1,\omega_p^mz_2,\omega_q^nz_3,\omega_q^nz_4)\quad
m,n\in{\bf Z},
\label{zpzq}
\end{eqnarray}
where $\omega_n=e^{2\pi i/n}$.
When $k\geq2$, we have an extra ${\bf Z}_k$ orbifolding
\begin{eqnarray}
(z_1,z_2,z_3,z_4)
\rightarrow (\omega_{kp}^mz_1,\omega_{kp}^mz_2,\omega_{kq}^{-m}z_3,\omega_{kq}^{-m}z_4)\quad
m\in{\bf Z}.
\label{zkaction}
\end{eqnarray}
Thus the background geometry is
\begin{equation}
{\cal M}_{p,q,k}=
({\bf C}^2/{\bf Z}_p
\times{\bf C}^2/{\bf Z}_q)/{\bf Z}_k.
\label{modpqk}
\end{equation}
When $N_I=N$, the Higgs branch moduli space is the
symmetric product of $N$ copies of the orbifold.

The rotational symmetry group of this manifold is
\begin{equation}
(SU(2)\times U(1))^2,
\label{isometry}
\end{equation}
and this agrees with the global symmetry of the Chern-Simons theory
shown in Table \ref{hyper.tbl}.
The R-symmetries
$SU(2)_A$ and $SU(2)_B$ act on ${\bf C}^2/{\bf Z}_p$ and ${\bf C}^2/{\bf Z}_q$,
respectively.

In order to obtain the orbifold above,
we should note that a certain subgroup of
$G_B$ is spontaneously broken by
the vacuum expectation value of the dual photon field ${\tilde a}$,
which is defined by
\begin{equation}
d {\tilde a}=\sum_{I=1}^n k_IA_I.
\end{equation}
Under the gauge symmetry
$A_I\rightarrow A_I+d\lambda_I$
the dual photon field is transformed by
\begin{equation}
{\tilde a}\rightarrow {\tilde a}+k_I\lambda_I.
\label{agauge}
\end{equation}
The dual photon field is periodic scalar field
with period $2\pi$ \cite{Martelli:2008si},
and the operator $e^{i\tilde a}$ carries $U(1)_I$ charge $k_I$.
The moduli space is parameterized by a set of mesonic operators.
We define mesonic operators as $G=G_{SU}\times G_B$ invariant operators.
By definition, they are neutral with respect to the baryonic symmetry $G_B$.
All trace operators are mesonic operators.
In addition to them, we can construct the following
mesonic operators
\begin{equation}
b=e^{-i{\tilde a}}\prod_{a=1}^p (h_a)^k,\quad
\tilde b=e^{i{\tilde a}}\prod_{\dot a=\dot 1}^{\dot q} (h_{\dot a})^k,
\label{bwtb}
\end{equation}
when $N=1$.

We suppress the R-symmetry indices in (\ref{bwtb}) .
The right hand side of the first equation in
(\ref{bwtb}) has $pk$ $SU(2)_A$ indices,
and we take symmetric part of these indices
to define $b$.
In terms of ${\cal N}=2$ language,
the two scalar field $h_I^1$ and $h_I^2$ are
chiral and anti-chiral fields, respectively,
and thus
$b^{1\cdots1}$ and $(b^{2\cdots2})^\dagger$ are chiral operators.
Due to the $SU(2)_A$ symmetry,
other components also
belong to certain short multiplets.
In the same way, $\tilde b$ has symmetric $qk$ $SU(2)_B$ indices.
When the size of the gauge groups is $N\geq2$,
we should replace the dual photon operators
in (\ref{bwtb})
by appropriate monopole operators \cite{zoo,Itzhaki:2002rc,Park:2008bk},
which have color indices needed to make
(\ref{bwtb}) gauge invariant.

If we put a large number of $N$ M2-branes at the tip of the orbifold
(\ref{modpqk}),
and take account of the back-reaction to the metric,
we obtain the dual geometry of this system.
It is
$AdS_4\times M_{p,q,k}$,
where $M_{p,q,k}$ is
the section of the orbifold
(\ref{orbi}) at
\begin{equation}
|z_1|^2+|z_2|^2+|z_3|^2+|z_4|^2=1.
\label{zzzz1}
\end{equation}
It is the following orbifold of seven-sphere.
\begin{equation}
M_{p,q,k}=({\bf S}^7/({\bf Z}_p\oplus{\bf Z}_q))/{\bf Z}_k.
\end{equation}
The radii of AdS$_4$ and $M_{p,q,k}$ are given by
\begin{equation}
R_{S^7}^6=(2R_{AdS_4})^6=\frac{(2\pi l_p)^6}{2\pi^4}Nkpq.
\label{radii}
\end{equation}
The radius of $M_{p,q,k}$ stands for that of the covering space
${\bf S}^7$.
The background metric is
\begin{equation}
ds^2
=R_{AdS_4}^2ds_{AdS_4}^2+R_{S^7}^2ds_{M_{p,q,k}}^2.
\label{metric}
\end{equation}

\section{Fractional D3-branes}\label{hw.sec}
As we mentioned in the last section,
the moduli space of an ${\cal N}=4$ Chern-Simons theory
depends only on the level $k$ and the numbers $p$ and $q$ of two
kinds of hypermultiplets.
It does not depend on the order of the two kinds of hypermultiplets in the
circular quiver diagrams.

This is quite similar to the situation
in the elliptic models of four-dimensional ${\cal N}=1$
supersymmetric gauge theories.
Such theories are generalizations of
Klebanov-Witten theory \cite{Klebanov:1998hh}, and
can be described by type IIA brane systems
which consist of D4-branes wrapped along ${\bf S}^1$
and NS5-branes intersecting with the D4-branes.
In this brane configuration,
NS5-branes are classified into two groups
according to their directions.
Let us call these NS5-branes with different directions
A-branes and B-branes.
On the D4-branes a four-dimensional gauge theory
which is described by a circular quiver diagram is realized.
If the number of A- and B-branes are $p$ and $q$,
the Coulomb branch moduli space of the gauge theory is
the symmetric product of a generalized conifold $uv=x^py^q$ \cite{Uranga:1998vf},
which depends only
on the numbers $p$ and $q$, and is independent of
the order of A- and B-branes along the ${\bf S}^1$.
The field theories sharing the same moduli space
are related by Seiberg duality \cite{Seiberg:1994pq},
and flow to the same effective theory
in the infra-red limit.
Such a relation between interchange of branes and Seiberg duality is
first pointed out in \citen{Elitzur:1997fh}.

\begin{figure}[htb]
\centerline{\includegraphics{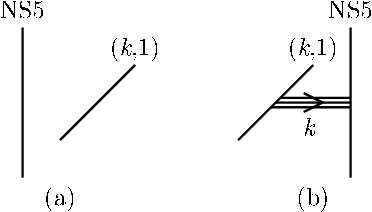}}
\caption{An example of D3-brane creation process is shown.
(a) is an initial configuration consisting of an NS5-brane and a $(k,1)$-fivebrane.
If $(k,1)$-brane is moved on the other side of the NS5-brane
as shown in (b),
$k$ D3-branes are created.}
\label{hw.eps}
\end{figure}

It is natural to expect that this is also the case for the three-dimensional
Chern-Simons theories we are considering here.
Such a brane exchange procedure is applied
to three-dimensional Chern-Simons theories in \citen{Giveon:2008zn,Niarchos:2008jb}.
Important difference of this duality and the four-dimensional one is
that in the three-dimensional case the brane exchange process in general generates
new D3-branes due to the Hanany-Witten effect \cite{Hanany:1996ie}.
(Figure \ref{hw.eps})

The purpose of this section is to classify theories which give the same moduli
space.
We assume that two theories realized by brane systems related by
a continuous deformation are dual to each other and flow to the same
infra-red fixed point.
We identify such theories, and the question considered here is
how much variety of inequivalent theories exist for given $p$, $q$, and $k$.
We study $k=1$ case first, and later discuss generalization to $k\geq 2$.

To realize $k=1$ theories, we use D3, NS5, and $(1,1)$ fivebranes.
The set of these three kinds of branes is equivalent to the set D3, NS5, and D5-branes
up to a certain $SL(2,{\bf Z})$ duality transformation.
Thus here we use the latter set of branes.
The directions of these branes are shown in Table \ref{setup.tbl}.
\begin{table}[htb]
\caption{The type IIB brane configuration is shown.}
\label{setup.tbl}
\begin{center}
\begin{tabular}{ccccccccccc}
\hline
\hline
& $0$ & $1$ & $2$ & $3$ & $4$ & $5$ & $6$ & $7$ & $8$ & $9$ \\
\hline
D3  & $\circ$ & $\circ$ & $\circ$ &&&& $\circ$ \\
NS5$\times p$ & $\circ$ & $\circ$ & $\circ$ &&&&& $\circ$ & $\circ$ & $\circ$ \\
D5$\times q$  & $\circ$ & $\circ$ & $\circ$ & $\circ$ & $\circ$ & $\circ$ \\
\hline
\end{tabular}
\end{center}
\end{table}
The D3-branes are wrapped on the compactified direction $x^6$.
The $p$ NS5-branes and $q$ D5-branes intersect with the D3-brane worldvolume and
divide the ${\bf S}^1$ into $n=p+q$ intervals.
We label the intersection points
by $I=1,2,\ldots,n$ in order along ${\bf S}^1$.
We emphasize that when we use $I$ as a label of fivebranes,
it represents the position of the fivebrane along ${\bf S}^1$.
In other words, $I$ is the label of slots in which we can put fivebranes.
We use $a,b=1,\ldots,p$ and $\dot a,\dot b=\dot1,\ldots,\dot q$
to label NS5-branes and D5-branes, respectively,
and the location of each fivebrane is specified by
giving one to one map $(a,\dot a)\rightarrow I$.
We call the choice of this map ``frame''.

We denote the number of D3-branes in the interval
between $I$-th and $I+1$-th fivebranes by $N_I$.
These numbers give the size of each $U(N)$ factor in
the gauge group in (\ref{gg}).
Let us define $m_I$ as the number of D3-branes emanating from
the $I$-th fivebrane.
By definition $N_I$ and $m_I$ are related by
\begin{equation}
m_I=N_I-N_{I-1}.
\end{equation}
Because $m_I$ are invariant under an overall shift
$N_I\rightarrow N_I+c$, ($c\in{\bf Z}$),
we cannot uniquely determine $N_I$ from $m_I$.
This degree of freedom represents integral D3-branes
wrapping around the whole ${\bf S}^1$.
We here focus only on the fractional brane charges
and use $m_I$ to represent D3-brane distributions.

The numbers $m_I$ in general change when the order of fivebranes
is changed by continuous deformations.
By this reason, to specify brane configuration,
we need not only to give $m_I$ but also to specify the frame,
the order of the fivebranes.
In the following we assume that we choose a particular frame.

With a fixed frame, a D3-brane distribution is specified by a vector
\begin{equation}
(m_1,m_2,\ldots,m_p|m_{\dot1},m_{\dot2},\ldots,m_{\dot q}).
\label{cv}
\end{equation}
We call this vector ``charge vector.''
By definition, the components of a charge vector must satisfy the constraint
\begin{equation}
\sum_{a=1}^pm_a
+\sum_{\dot a=\dot 1}^{\dot q}m_{\dot a}
=0.
\label{summk}
\end{equation}
The set of charge vectors,
whose components are constrained by
(\ref{summk}), forms the group
\begin{equation}
\Gamma={\bf Z}^{p+q-1}.
\end{equation}
We should not regard the group $\Gamma$ as the group characterizing the
conserved charge of fractional D3-branes because D3-brane distributions corresponding
to different elements of $\Gamma$ may be continuously deformed to one another.
We should regard charges of such brane configurations
as the same.

Let us move NS5-brane $a$ in the positive direction
along ${\bf S}^1$ until it comes back to the original position.
This process does not change the frame, but changes the charge vector.
When the NS5-brane passes through an D5-brane $\dot b$,
$m_a$ decreases by one and $m_{\dot b}$ increases by one.
(Now we assume $k=1$.)
When the NS5-brane comes back to the original position,
the charge vector changes by
\begin{eqnarray}
{\bf v}_a=(0,\ldots,-q,\ldots,0|1,\ldots,1)
   =-q{\bf e}_a+\sum_{\dot b=\dot 1}^{\dot q}{\bf e}_{\dot b}\in \Gamma,
\label{def1}
\end{eqnarray}
where ${\bf e}_a$ (${\bf e}_{\dot b}$) is the unit vectors
whose $a$-th ($\dot b$-th) component is $1$.
Note that ${\bf e}_a$ and ${\bf e}_{\dot b}$
themselves are not elements of $\Gamma$
because they do not satisfy the constraint (\ref{summk}).
Similarly, if we move a D5-brane $\dot a$ around the ${\bf S}^1$,
the charge vector changes by
\begin{eqnarray}
{\bf w}_{\dot a}=(1,\ldots,1|0,\ldots,-p,\ldots,0)
   =\sum_{b=1}^p{\bf e}_b-p{\bf e}_{\dot a}\in \Gamma.
\label{def2}
\end{eqnarray}
When we identify configurations deformed by continuous deformation to one another,
these vectors should be identified with $0$.
Therefore, the group describing the charge of fractional branes
is the quotient group $\Gamma/H$
where $H$ is the subgroup of $\Gamma$ generated by the
vectors ${\bf v}_a$ and ${\bf w}_{\dot b}$.
This is given by
\begin{equation}
\Gamma/H=
({\bf Z}_p^{q-1}\oplus
{\bf Z}_q^{p-1}\oplus
{\bf Z}_{pq})/
({\bf Z}_p\oplus{\bf Z}_q).
\label{quotirnt}
\end{equation}
In the rest of this section, we explain how this expression of the
quotient group is obtained.

As we mentioned above, the $p+q$ vectors ${\bf e}_a$ and ${\bf e}_{\dot a}$
are not elements of $\Gamma$.
Let us choose $p+q-1$ linearly independent basis in $\Gamma$.
We take the following vectors.
\begin{eqnarray}
{\bf f}_a&=&{\bf e}_a-{\bf e}_p
\quad (a=1,\ldots, p-1),\\
{\bf g}_{\dot a}&=&{\bf e}_{\dot a}-{\bf e}_{\dot q}
\quad (\dot a=\dot1,\ldots, \dot{(q-1)}),\\
{\bf h}&=&{\bf e}_p-{\bf e}_{\dot q}.
\end{eqnarray}
We can easily check that these vectors span $\Gamma$.
In order to obtain (\ref{quotirnt}),
we define the subgroup $H'\subset H$ generated by
the following elements in $H$.
\begin{eqnarray}
{\bf v}_p-{\bf v}_a&=&q{\bf f}_a\quad(a=1,\ldots,p-1),\\
{\bf w}_{\dot q}-{\bf w}_{\dot a}&=&p{\bf g}_{\dot a}\quad(\dot a=\dot 1,\ldots,\dot{(q-1)}),\\
-p{\bf v}_p+q{\bf w}_{\dot q}-\sum_{\dot b=\dot 1}^{\dot q}{\bf w}_{\dot b}
&=&pq{\bf h}.
\end{eqnarray}
We can easily show that
\begin{equation}
\Gamma/H'=
{\bf Z}_p^{q-1}\oplus
{\bf Z}_q^{p-1}\oplus
{\bf Z}_{pq},\quad
H/H'={\bf Z}_p\oplus{\bf Z}_q,
\end{equation}
and the relation $\Gamma/H=(\Gamma/H')/(H/H')$
gives (\ref{quotirnt}).

\section{Fractional M2-branes}\label{frac.sec}
In this section we reproduce the quotient group
(\ref{quotirnt}) as the $3$-cycle homology of the internal space $M_{p,q}:=M_{p,q,1}$
of the dual geometry for $k=1$.

Let us remember how the geometry is obtained as a dual configuration
from the IIB brane system in Table \ref{setup.tbl}.
We first perform T-duality transformation
along $x^6$ and then lift the system to M-theory configuration.
As the result we have the configuration shown in Table \ref{geom.tbl}.
\begin{table}[htb]
\caption{The dual M-theory geometry is shown. 
Shrinking cycles  are denoted by  ``s.''
}
\label{geom.tbl}
\begin{center}
\begin{tabular}{cccccccccccc}
\hline
\hline
& $0$ & $1$ & $2$ & $3$ & $4$ & $5$ & $6$ & $7$ & $8$ & $9$ & $M$ \\
\hline
M2-branes & $\circ$ & $\circ$ & $\circ$ \\
$q$ KK monopoles & $\circ$ & $\circ$ & $\circ$ & &&& s & $\circ$ & $\circ$ & $\circ$ & $\circ$ \\
$p$ KK monopoles & $\circ$ & $\circ$ & $\circ$ & $\circ$ & $\circ$ & $\circ$ & $\circ$ &&&& s \\
\hline
\end{tabular}
\end{center}
\end{table}
Two kinds of fivebranes are mapped to purely geometric objects,
Kaluza-Klein (KK) monopoles.
In general, $Q$ coincident KK monopoles are described as an $A_{Q-1}$-type orbifold,
and the geometry shown in Table \ref{geom.tbl} is the product of $A_{p-1}$ and $A_{q-1}$ singularities.
This is nothing but the orbifold ${\cal M}_{p,q}$ in (\ref{orbi}).

Fractional M2-branes are realized
as M5-branes wrapped on $3$-cycles in $M_{p,q}$.
The homologies $H_i(M_{p,q},{\bf Z})$ are given by%
\footnote{We obtained these homologies by careful search for cycles in the manifold.
We have checked their consistency with the Poincare duality and the Mayer-Vietoris exact sequence.}
\begin{eqnarray}
&&
H_0={\bf Z},\quad
H_1=0,\quad
H_2={\bf Z}^{p+q-2},\quad
H_3=({\bf Z}_p^{q-1}\oplus{\bf Z}_q^{p-1}\oplus{\bf Z}_{pq})/({\bf Z}_{p}\oplus{\bf Z}_q),
\nonumber\\&&
H_4=0,\quad
H_5={\bf Z}^{p+q-2},\quad
H_6=0,\quad
H_7={\bf Z}.
\label{homology}
\end{eqnarray}
The relevant homology $H_3(M_{p,q},{\bf Z})$ is pure torsion,
and it coincides with the group of fractional D3-branes
studied in the previous section.
In the following
we explain how the $H_3$ group in (\ref{homology}) is obtained.

\begin{figure}[htb]
\centerline{\includegraphics{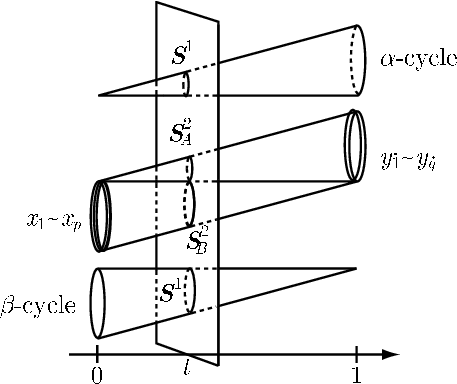}}
\caption{The orbifold is represented as a fibration over the segment $0\leq t\leq1$.}
\label{s7section.eps}
\end{figure}

For the purpose of considering cycles in $M_{p,q}$,
it is convenient to
represent $M_{p,q}$ as a ${\bf T}^2$ fibration over $B={\bf S}^5$.
This fibration is defined in the following way.
We introduce a real coordinate $0\leq t\leq 1$ by
rewriting (\ref{zzzz1}) as
\begin{equation}
|z_1|^2+|z_2|^2=t,\quad
|z_3|^2+|z_4|^2=1-t.
\label{coordt}
\end{equation}
At a generic value of $t$, this defines two $3$-spheres,
and the orbifold action (\ref{zpzq}) makes them Lens spaces $L_p$ and $L_q$.
The manifold $M_{p,q}$ is represented as $L_p\times L_q$ fibration
over the segment $0\leq t\leq 1$.
Each of Lens spaces $L_p$ and $L_q$ can be represented
as ${\bf S}^1$ fibration over
$2$-sphere.
For $L_p$, which is rotated by the $SU(2)_A$ R-symmetry,
we refer to the base manifold and the fiber as ${\bf S}_A^2$ and
$\alpha$-cycle, respectively.
We also define ${\bf S}_B^2$ and $\beta$-cycle for the other Lens space
$L_q$, which is rotated by $SU(2)_B$.
(Figure \ref{s7section.eps})
Due to the ${\bf Z}_p\oplus {\bf Z}_q$ orbifolding,
the periods of $\alpha$ and $\beta$-cycles ara
$2\pi/p$ and $2\pi/q$, respectively.
If we combine ${\bf S}_A^2$, ${\bf S}_B^2$, and the segment parameterized by $t$,
they form a $5$-sphere $B={\bf S}^5$.
We can regard the orbifold $M_{p,q}$ as a ${\bf T}^2$ fibration over $B$.

\begin{figure}[htb]
\centerline{\includegraphics{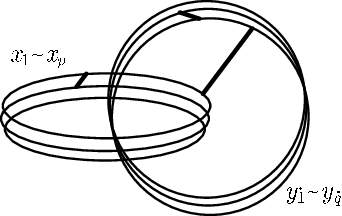}}
\caption{The three segments connecting cycles are examples of
three types of three-cycles in the orbifold.}
\label{3cycles.eps}
\end{figure}

At $t=0$, which defines ${\bf S}^2\subset B$,
the Lens space $L_p$ shrinks and so does the $\alpha$-cycle.
Similarly, on ${\bf S}^2\subset B$ with $t=1$
the $\beta$-cycle shrinks.
These ${\bf S}^2$ link to each other in $B$.
By blowing up the singularities,
these ${\bf S}^2$s split into $p$ and $q$ ${\bf S}^2$s, respectively.%
\footnote{We blow-up the singularities only to make cycles well-defined.
When we compute the volume of five-cycles later, we consider
the orbifold limit.}
We call them $x_a$ ($a=1,\ldots,p$) and $y_{\dot a}$ ($\dot a=\dot 1,\ldots,\dot q$).
(Figure \ref{3cycles.eps})
We can follow the IIB/M duality to see that each of them
corresponds to the fivebrane with the same index.

$3$-cycles in $M_{p,q}$ can be represented as ${\bf T}^2$ fibrations
over segments in the base manifold $B={\bf S}^5$.
There are three types of segments connecting two loci of degenerate fiber.
(Figure \ref{3cycles.eps})
We denote a segment connecting a point in $x_a$ and a point in $x_b$
by $[x_a,x_b]$.
We similarly define $[y_{\dot a},y_{\dot b}]$ and $[x_a,y_{\dot b}]$.
We also adopt the notation
\begin{equation}
S^\alpha,
S^\beta,
S^{\alpha\beta}\subset M_{p,q}
\end{equation}
for the manifold obtained by combining
a subset $S\subset B$ and fibers indicated as superscripts.
$S^{\alpha\beta}$ is the ${\bf T^2}$ fibration over $S$.
$S^{\alpha\beta}$ can be regarded as $\beta$-fibration
over a certain base manifold, which is isomorphic to $S^\alpha$.
$S^\alpha$ is a global section in this fiber bundle.
Therefore, $S^\alpha$ can be defined only when $\beta$-cycle fibration
over $S$ has a global section.
Similarly, we can define $S^\beta$ when $\alpha$-cycle fiber has
trivial topology over $S$.

With these notations, we can represent $3$-cycles
generating $H_3$ as
\begin{equation}
[x_a,x_b]^{\alpha\beta},\quad
[y_{\dot a},y_{\dot b}]^{\alpha\beta},\quad
[x_a,y_{\dot b}]^{\alpha\beta}.
\label{threekin}
\end{equation}
Because one of $\alpha$ and $\beta$ cycles shrinks at the endpoints of
the segments, these are closed $3$-cycles.
The topology of $[x_a,x_b]^{\alpha\beta}$
and $[y_{\dot a},y_{\dot b}]^{\alpha\beta}$ is ${\bf S}^2\times{\bf S}^1$,
and that of $[x_a,y_{\dot b}]^{\alpha\beta}$ is ${\bf S}^3$.

\begin{figure}[htb]
\centerline{\includegraphics{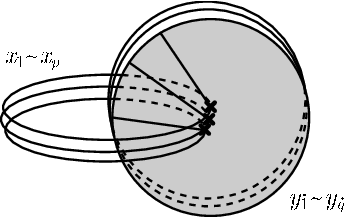}}
\caption{The $\beta$-cycle fibration over the gray disk
with the segments removed is
an example of unwrapping four-chains.}
\label{unwrap.eps}
\end{figure}

These $3$-cycles are not linearly independent.
There are combinations of cycles which can be unwrapped.
Let us consider
\begin{equation}
\sum_{a=1}^p[x_a,y_{\dot b}]^{\alpha\beta}.
\label{sumab}
\end{equation}
This union of $3$-cycles can be unwrapped in $M_{p,q}$.
This can be shown by giving a $4$-chain whose boundary is (\ref{sumab}).
Such an ``unwrapping chain'' is constructed in the following way.
Because $\pi_2({\bf S}^5)=0$,
there is a three dimensional disk ${\bf D}^3\subset B$ whose boundary is $y_{\dot b}$.
(The gray disk in Figure \ref{unwrap.eps})
We call this $Y_{\dot b}$.
(We also define $X_a$ in the same way for $x_a$.)
This disk intersects once with every $x_a$ ($a=1,\ldots,p$).
Let $\bar Y_{\dot b}$ be the subset of $Y_{\dot b}$ obtained
by removing segments connecting these intersecting points and $y_{\dot b}$
(the segments in Fig \ref{unwrap.eps}) from the disk.
\begin{equation}
\bar Y_{\dot b}=Y_{\dot b}\backslash\sum_{a=1}^p[x_a,y_{\dot b}].
\end{equation}
Because $\bar Y_{\dot b}$ is contractible,
we can define $\bar Y_{\dot b}^\beta$.
(Note that we cannot define $Y_{\dot b}^\beta$ because the $\alpha$ cycle is
twisted around the intersecting points of $Y_{\dot b}$ and $x_a$.)
We can see that the boundary of the manifold $\bar Y_{\dot b}^\beta$ is
\begin{equation}
\partial \bar Y_{\dot b}^\beta
=
\sum_{a=1}^p[x_a,y_{\dot b}]^{\alpha\beta}.
\label{dy}
\end{equation}
This may seem at first sight strange because
although $\bar Y_{\dot b}^\beta$ does not wrap the $\alpha$-cycle
its boundary does.
Let us explain this situation by taking Hopf fibration of ${\bf S}^3$
as a simple example. 
By the Hopf fibration ${\bf S}^3$ is described as the
${\bf S}^1$ fibration over ${\bf S}^2$.
Let $(\theta,\phi)$ be the polar coordinates of
the base ${\bf S}^2$.
The first Chern class of this fiber bundle is $1$,
so that we cannot globally define the coordinate
of the fiber.
We cover the base ${\bf S}^2$ bytwo patches,
north patch ($0\leq\theta<\pi$) and south patch
($0<\theta\leq\pi$),
and define fiber coordinate $0\leq \psi\leq 2\pi$
separately in each patch.
Let $\psi_N$ and $\psi_S$ be that in north and south patch,
respectively.
These two coordinats are paseted by the relation $\psi_N=\psi_S+\phi$.
Due to the non-vanishing 
first Chern class,
we cannot take a global section in this fiber bundle.
In order to define sections, we need to remove at least one
point from the base ${\bf S}^2$.
Let us take south patch.
We can define, for example, the section
\begin{equation}
0<\theta\leq\pi,\quad
0\leq\phi<2\pi,\quad
\psi_S=0.
\end{equation} 
At the boundary $\theta=0$ of south patch, the north pole,
this section wrap the fiber ${\bf S}^1$.
This becomes obvious if we use the coordinate $\psi_N$,
which includes the north pole.
The boundary is given by
\begin{equation}
\theta=\pi,\quad
0\leq\phi<2\pi,\quad
\psi_N=\phi.
\end{equation}
This winds once along the fiber ${\bf S}^1$.
This result makes sense from the fact that the homology 
$H_1({\bf S}^3)$ vanishes.
Any $1$-cycle on ${\bf S}^3$ can be unwrapped and
represented as the boundary of a $2$-chain.



By exchanging the role of $X_a$ and $Y_{\dot b}$, we can also show
\begin{equation}
\partial\bar X_a^\alpha=
\sum_{\dot b=1}^{\dot q}[x_a,y_{\dot b}]^{\alpha\beta}.
\label{sumab2}
\end{equation}

From (\ref{dy}) and (\ref{sumab2}),
we obtain the homology relation
\begin{equation}
\sum_{a=1}^p[x_a,y_{\dot b}]^{\alpha\beta}
=\sum_{{\dot b}=\dot 1}^{\dot q}[x_a,y_{\dot b}]^{\alpha\beta}=0.
\label{twobdr}
\end{equation}

To clarify the relation between the IIB picture and the M-theory picture,
we define formal basis ${\bf x}_a$ and ${\bf y}_{\dot b}$
and rewrite cycles as
$[x_a,x_b]^{\alpha\beta}={\bf x}_a-{\bf x}_b$ and so on.
A general superposition of cycles,
which is depicted as a junction in $B$, can be written as a
linear combination
\begin{equation}
{\bf j}=\sum_{a=1}^pm_a{\bf x}_a+
\sum_{\dot a=\dot 1}^{\dot q}m_{\dot a}{\bf y}_{\dot a},
\label{junc}
\end{equation}
where the coefficients must satisfy the constraint (\ref{summk}).
We have one to one correspondence between $3$-cycles in $M_{p,q}$ and
D3-brane distributions in IIB picture by simply identifying the coefficients
in (\ref{junc}) to the components of the charge vector (\ref{cv}).
Via this isomorphism the boundaries (\ref{dy}) and (\ref{sumab2})
correspond to ${\bf w}_{\dot b}$ and ${\bf v}_a$,
the generators of $H$,
and the relation (\ref{twobdr}) defines the homology $H_3$ as the same coset
group $\Gamma/H$ in (\ref{quotirnt}).

\section{Five cycles and baryonic operators}\label{baryon.sec}
In this section, we discuss the relation between M5-branes wrapped on
five cycles and baryonic operators in the ${\cal N}=4$ Chern-Simons theory.
In the case of ABJM model, such analysis is done in \citen{Park:2008bk},
and the conformal dimension and multiplicity
of baryonic operators are reproduced on the gravity side.
We extend the results to ${\cal N}=4$ Chern-Simons theories.

As in the previous section, we consider $k=1$ case.
As is given in (\ref{homology}),
the five-cycle homology of $M_{p,q}$ is
\begin{equation}
H_5(M_{p,q},{\bf Z})={\bf Z}^{p+q-2},
\end{equation}
and when $p+q\geq3$, there exist non-trivial cycles
on which M5-branes can be wrapped.
If we represent $M_{p,q}$ as the ${\bf T}^2$ fibration over $B={\bf S}^5$,
the five-cycles can be written as the ${\bf T}^2$ fibrations over
three-disks.
\begin{equation}
\Omega_a:=X_a ^{\alpha\beta},\quad
\Omega_{\dot a}:=Y_{\dot a} ^{\alpha\beta}.
\label{omegadef}
\end{equation}
These generate the homology $H_5(M_{p,q},{\bf Z})$.

The number of the cycles in (\ref{omegadef}) is larger than the dimension
of $H_5(M_{p,q},{\bf Z})$ by two, and there should be two relations among the cycles in (\ref{omegadef}).
Indeed, we have the following homology relations
\begin{equation}
\sum_{a=1}^p \Omega_a=
\sum_{\dot{a}=\dot{1}}^{\dot{q}} \Omega_{\dot{a}}=0.
\label{XYtrivial}
\end{equation}
As we have done in section \S\ref{frac.sec} for three-cycles,
we can give these linear combinations as the boundaries of
unwrapping $6$-chains.
We define 
a submanifold $\bar B\subset B$ by
\begin{equation}
\bar{B} = B \backslash\left(
\sum_{a=1}^p X_a+
\sum_{\dot a=\dot 1}^{\dot p} Y_{\dot a}\right).
\end{equation}
If we could draw ${\bf S}^2$ enclosing $x_a$ in $\bar B$,
the $\alpha$-cycle fiber would have
non-trivial twist on the ${\bf S}^2$.
However, such ${\bf S}^2$ do not exist in $\bar B$
because we removed the disks $X_a$.
Thus the $\alpha$-cycle fiber over $\bar B$ has trivial topology
and we can define global sections.
Similarly, thanks to the removal of $Y_{\dot a}$,
the $\beta$-cycle fiber also have the trivial topology.
Because there is a global section associated with
$\alpha$-cycle over $\bar B$,
the manifold $\bar{B}^{\beta}$ is well-defined,
and its boundary is
\begin{equation}
\partial \bar B^\beta=\sum_{a=1}^p X_a ^{\alpha \beta}.
\end{equation}
We also obtain
\begin{equation}\label{Ytrivial}
\partial\bar B^\alpha
=\sum_{\dot a=1}^{\dot q} Y_{\dot a}^{\alpha\beta}.
\end{equation}
As a result, we obtain the
relations (\ref{XYtrivial}).

What are the corresponding baryonic operators
on the gauge theory side?
A natural guess is that these five-cycles are dual to
the following operators in the Chern-Simons theory:
\begin{eqnarray}
\Omega_a&\leftrightarrow&
B_a^{\alpha_1\alpha_2\cdots\alpha_N}
=\epsilon_{i_1 \cdots i_N}
\epsilon^{j_1 \cdots j_N}
h_a^{\alpha_1}{}^{i_1}_{j_1}
\cdots
h_a^{\alpha_N}{}^{i_N}_{j_N},
\label{baryonic1}\\
\Omega_{\dot a}&\leftrightarrow&
B_{\dot a}^{\dot\alpha_1\dot\alpha_2\cdots\dot\alpha_N}
=\epsilon_{i_1 \cdots i_N}
\epsilon^{j_1 \cdots j_N}
h_{\dot a}^{\dot\alpha_1}{}^{i_1}_{j_1}
\cdots
h_{\dot a}^{\dot\alpha_N}{}^{i_N}_{j_N}.
\label{baryonic2}
\end{eqnarray}
Each of these operators is charged under the baryonic symmetry $G_B$,
and cannot be decomposed into mesonic operators,
which are $G_B$ neutral.
However, the products
\begin{equation}
\prod_{a=1}^pB_a,\quad
\prod_{\dot a=1}^{\dot q}B_{\dot a}
\end{equation}
carry the same baryonic charge as $e^{iN{\tilde a}}$ and $e^{-iN{\tilde a}}$, respectively,
and by multiplying appropriate power of operator $e^{i{\tilde a}}$,
we can construct neutral operators with respect to the baryonic symmetries.
This strongly suggests that these can be decomposable to
the mesonic operators as
\begin{equation}
e^{-i N {\tilde a}}  \prod_{a=1}^pB_a\sim b^N,\quad
e^{i N{\tilde a}} \prod_{\dot a=1}^{\dot q}B_{\dot a}\sim\tilde b^N.
\end{equation}
This decomposability corresponds to the homology relation
(\ref{XYtrivial}) among the five cycles.

\begin{figure}[htb]
\centerline{\includegraphics{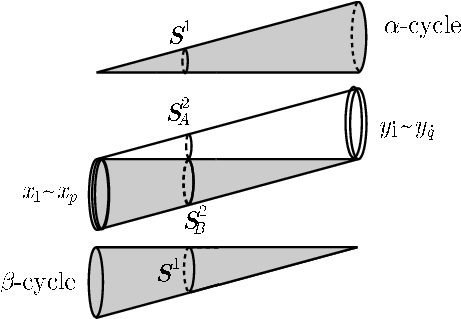}}
\caption{The shaded region is a nontrivial 5-cycle $X _{a} ^{\alpha \beta}$. }
\label{5-cycle.eps}
\end{figure}

As a non-trivial check of the duality,
let us compare the mass of the wrapped M5-branes
and the conformal dimension of the operators.
According to the standard AdS/CFT dictionary,
the conformal dimension $\Delta$ of an operator
and the mass $M$ of the corresponding object are related by
$\Delta=R_{{\rm AdS}_4}M$.
In the case of an M5-brane wrapped on $\Omega_a$,
this relation becomes
\begin{equation}
\Delta
=R_{AdS_4}T_{M5} R_{S^7}^5\Vol(\Omega_a)
=\frac{Npq}{2\pi^3}\Vol(\Omega_a).
\label{deltais}
\end{equation}
where $\Vol(\Omega_a)$ is the volume of the 5-cycle $\Omega_a$
in $M_{p,q}$ with radius $1$,
and to obtain the last expression we used
(\ref{radii}) with $k=1$ and the M5-brane tension $T_{M5}=2\pi/(2\pi l_p)^6$.
Let us calculate the volume of the 5-cycle.
The $5$-cycle $\Omega_a$,
which is represented as a fiber bundle over the segment $0\leq t\leq 1$,
is illustrated as the shaded region in Figure \ref{5-cycle.eps}.
The radii of two $3$-spheres
defined by (\ref{coordt}) are
$r_1=t^{1/2}$ and $r_2=(1-t)^{1/2}$, respectively.
The cross-section at $t$ is
${\bf S}^1\times{\bf S}^2\times{\bf S}^1$
with their radii $r_1/p$, $r_2/2$, $r_2/q$, respectively\footnote{%
It is known that when a unit ${\bf S}^3$ is represented
by the ${\bf S}^1$ fibration over ${\bf S}^2$, 
the radii of ${\bf S}^1$ and ${\bf S}^2$ are
$1$ and $1/2$ respectively.}. 
Hence the volume of the 5-cycle is 
\begin{equation}
\Vol(\Omega_a)
=\int_{t=0}^{t=1}ds
\left(\frac{2 \pi r_1}{p}\right)\times
\left(4\pi\left(\frac{r_2}{2}\right)^2\right)\times
\left(\frac{2\pi r_2}{q}\right)
=\frac{\pi^3}{pq},
\label{volume}
\end{equation}
where $ds$ is the line element with respect to the parameter $t$ computed as
\begin{equation}
ds^2=dr_1^2+dr_2^2=\frac{1}{4t(1-t)}dt^2.
\end{equation}
(The volume (\ref{volume})
is simply $\Vol({\bf S}^5)/pq$
because
the five-cycles considered here are orbifolds of large ${\bf S}^5$ in ${\bf S}^7$.)
We obtain the same result for $5$-cycles $Y_{\dot a}^{\alpha \beta}$.
By substituting this into
(\ref{deltais}) we obtain
\begin{eqnarray}
\Delta=\frac{1}{2}N,
\label{deltaeqn2}
\end{eqnarray}
and this agrees with the conformal dimension of the baryonic operators
(\ref{baryonic1}) and (\ref{baryonic2}).
(\ref{deltaeqn2}) is consistent with the result of more general analysis in \citen{Yee:2006ba}
for generic toric tri-Sasakian manifolds.

The degeneracy of the baryonic operators are explained
in the same way as the Klebanov-Witten theory \cite{Berenstein:2002ke}.
The collective coordinates of five-cycle $\Omega_a$
are the coordinates in the transverse direction
${\bf S}^2_A$, on which $SU(2)_A$ acts as rotation.
The seven-form flux in the background plays
a role of magnetic field on ${\bf S}_A^2$ and the amount of the flux is $N$.
Therefore, the effective theory of the corrective coordinates
is the theory of a charged particle in ${\bf S}_A^2$ with
$N$ unit magnetic flux.
The ground states of the particle are the $N+1$ states
at the lowest Landau level \cite{Coleman:1982cx}
belonging to the spin $N/2$ representation of $SU(2)_A$.
This degeneracy agrees with that of the baryonic operators $B_a$.
In the same way, we can explain the degeneracy of $B_{\dot a}$ 
as that of the lowest Landau level of a charged particle
in the transverse direction ${\bf S}^2_B$.

\section{Generalization to $k\geq2$}\label{kgeq2.sec}
In this section we generalize the analysis in the
previous sections to the case of $k\geq2$.

Let us first consider fractional D3-brane charges in the type IIB brane setup.
We can realize the Chern-Simons theory at level $k$ by replacing
D5-branes by $(k,1)$ fivebranes.
We can again represent distributions of D3-branes
by charge vectors (\ref{cv}) with their components constrained by (\ref{summk}).
The only difference from the $k=1$ case is that when a $(k,1)$-brane
and an NS5-brane
pass through each other, not one but $k$ D3-branes are generated.
As a result,
the vectors (\ref{def1}) and (\ref{def2}) are
multiplied by the extra factor $k$.
Namely, we should replace the subgroup $H$ by $kH$
which is generated by
$k{\bf v}_a$ and $k{\bf w}_{\dot a}$,
and the quotient group becomes
\begin{equation}
\Gamma/(kH)
=(\Gamma/kH')/(H/H')
=({\bf Z}_{kq}^{p-1}\oplus{\bf Z}_{kp}^{q-1}\oplus{\bf Z}_{kpq})/({\bf Z}_p\oplus{\bf Z}_q).
\label{h3mpqk}
\end{equation}

On the other hand,
the homologies $H_i(M_{p,q,k},{\bf Z})$ are
\begin{eqnarray}
&&
H_0={\bf Z},\quad
H_1={\bf Z}_k,\quad
H_2={\bf Z}^{p+q-2},\quad
H_3=({\bf Z}_{kp}^{q-1}\oplus{\bf Z}_{kq}^{p-1}\oplus{\bf Z}_{kpq})/({\bf Z}_{p}\oplus{\bf Z}_q),
\nonumber\\&&
H_4=0,\quad
H_5={\bf Z}^{p+q-2}\oplus{\bf Z}_k,\quad
H_6=0,\quad
H_7={\bf Z}.
\label{homology2}
\end{eqnarray}
We find that the homology $H_3$ is again identical to (\ref{h3mpqk}).
Let us construct the homology $H_3(M_{p,q,k},{\bf Z})$ more explicitly.
When the level $k$ is greater than $1$, we have additional
${\bf Z}_k$ factor in the orbifold group.
As is shown in (\ref{zkaction}),
the generator of ${\bf Z}_k$ shifts both $\alpha$ and $\beta$ cycle by $1/k$
of their periods.
Because two cycles nowhere shrink at the same time,
this action does not generate fixed points.
The ${\bf Z}_k$ identification in the ${\bf T}^2$ fiber generates new cycles,
which are not integral linear combinations of $\alpha$ and $\beta$.
They are multiples of
\begin{equation}
\gamma=\frac{1}{k}(\alpha-\beta).
\end{equation}
As a result, the $2$-cycle defined as the product of $\alpha$ and $\beta$ is
not the fundamental ${\bf T}^2$ but its multiple $k{\bf T}^2$.
Thus the cycles (\ref{threekin})
are decomposed into
$k$ copies of the following elementary cycles.
\begin{equation}
[x_a,x_b]^{\alpha\gamma},\quad
[y_{\dot a},y_{\dot b}]^{\alpha\gamma},\quad
[x_a,y_{\dot b}]^{\alpha\gamma}.
\label{threekin2}
\end{equation}
Due to this fact, the boundary of unwrapping $4$-chains (\ref{sumab2})
and (\ref{dy}) are replaced by
\begin{equation}
\partial\bar X_a^\alpha
=k(-q{\bf x}_a+\sum_{\dot b=\dot1}^{\dot q}{\bf y}_{\dot b}),\quad
\partial\bar Y_{\dot a}^\beta
=k(\sum_{b=1}^p{\bf x}_b-p{\bf y}_{\dot a}),
\end{equation}
where we defined the formal basis ${\bf x}_a$ and ${\bf y}_{\dot a}$
by $[x_a,x_b]^{\alpha\gamma}={\bf x}_a-{\bf x}_b$ and so on.
These precisely correspond to the vectors $k{\bf v}_a$ and $k{\bf w}_{\dot a}$,
and thus the homology $H_3$ becomes isomorphic to the quotient $\Gamma/kH$
in (\ref{h3mpqk}).

Next, let us consider the relation between baryonic operators and
$5$-cycle homology $H_5$ for $k\geq2$.
In the case of $k\geq2$, the generators of the homology
in (\ref{omegadef}) should be replaced by
\begin{equation}
\Omega_a=X_a^{\alpha\gamma},\quad
\Omega_{\dot a}=Y_{\dot a}^{\alpha\gamma},
\label{omegak}
\end{equation}
and M5-branes wrapped on these generating cycles are identified with
the baryonic operators $B_a$ and $B_{\dot a}$.
We can again easily check that the volume of the five-cycles
correctly reproduce the conformal dimension $\Delta=N/2$.
The $p+q$ generators (\ref{omegak}) are not linearly independent,
and
we can take $\bar B^\alpha$, $\bar B^\beta$,
and $\bar B^\gamma$
as unwrapping $6$-chains which give the relation among these
generators.
Their boundaries are
\begin{eqnarray}
\partial\bar B^\alpha
&=&\sum_{\dot a=\dot 1}^{\dot q} Y_{\dot a}^{\alpha\beta}
=k\sum_{\dot a=\dot 1}^{\dot q} \Omega_{\dot a},\label{dbalpha}\\
\partial\bar B^\beta
&=&\sum_{a=1}^p X_a^{\alpha\beta}
=k\sum_{a=1}^p \Omega_a,\label{dbbeta}\\
\partial\bar B^\gamma
&=&\sum_{a=1}^p X_a^{\gamma\beta}+\sum_{\dot a=\dot 1}^{\dot q} Y_{\dot a}^{\gamma\alpha}
=\sum_{a=1}^p \Omega_a+\sum_{\dot a=\dot 1}^{\dot q} \Omega_{\dot a}.
\label{dbgamma}
\end{eqnarray}
Namely, these linear combinations of five-cycles are in trivial element
of the homology $H_5$.
By dividing the group ${\bf Z}^{p+q}$ generated by the $p+q$ basis
$\Omega_a$ and $\Omega_{\dot a}$ by
its subgroup ${\bf Z}^2$ generated by the above boundaries,
we obtain the $H_5$ homology in (\ref{homology2}).

On the field theory side,
the linear dependence of the five-cycles are interpreted as the
decomposability of the products of the baryonic operators into the
mesonic operators.
The first two, (\ref{dbalpha}) and (\ref{dbbeta}),
correspond to the product of $B_a$ and $B_{\dot a}$, respectively, and
are decomposed into $N$-th power of operators defined in (\ref{bwtb}).
\begin{equation}
e^{-iN{\tilde a}}\prod_{a=1}^pB_a^k\sim b^N,\quad
e^{iN{\tilde a}}\prod_{\dot a=1}^{\dot q}B_{\dot a}^k\sim \tilde b^N,
\end{equation}
The third boundary (\ref{dbgamma})
corresponds to the product of all the $p+q$ baryonic operators,
and it can be decomposed into trace operators.
\begin{equation}
\prod_{a=1}^pB_a^k\prod_{\dot a=1}^{\dot q}B_{\dot a}^k
\sim\left(\tr(\prod_{a=1}^p h_a\prod_{\dot a=\dot1}^{\dot q}h_{\dot a})\right)^N.
\end{equation}

The degeneracy of baryonic operators for $p+q\geq3$ are again reproduced
in the same way as the $k=1$ case.
In the case of $p=q=1$ (ABJM model), we need a special treatment
because the global symmetry (\ref{isometry}) is enhanced to
$SU(4)\times U(1)$ and the motion of collective coordinates
are treated as a point particle in $SU(4)/(SU(3)\times U(1))$.
This is considered in \citen{Park:2008bk} and
the correct multiplicity is obtained.

\section{Quark-baryon transition}\label{qb.sec}
In \S\ref{baryon.sec}, we studied the relation between
wrapped M5-branes and
baryonic operators $B_I$.
We can relate them more directly by using IIB/M duality
explained in \S\ref{frac.sec}.
By following the duality, we can easily see that
an M5-brane wrapped on $\Omega_I$
is dual to a D3-brane disk ending on fivebrane $I$,
and as we explain below, 
the D3-brane disk can be continuously deformed to $N$ open strings
corresponding to the constituent bi-fundamental quarks.
(Similar transition in different brane systems are also
considered in \citen{Imamura:2006ie,Lee:2006hw}.)

Before we explain the deformation, we comment on
a relevant fact about flux conservation
on the worldvolume of a D3-brane ending on an NS5-brane.
The $U(1)$ gauge field $A$ on an NS5-brane
electrically couples to endpoints of D-strings
on the NS5-brane.
This is the case, too, for magnetic flux $f=da$ on D3-branes,
which can be regarded as D-strings dissolved in
the D3-brane worldvolume.
This coupling is described as the action
\begin{equation}
S=\frac{1}{2\pi}\oint_{\partial D3} A\wedge f.
\end{equation}
By integrating by part, this is rewritten as
\begin{equation}
S=\frac{1}{2\pi}\oint_{\partial D3} a\wedge F,
\end{equation}
and this implies that the flux $F=dA$ on the NS5-brane
behaves as an electric charge on the boundary of the D3-brane
coupled by the gauge field $a$.
If the D3-brane worldvolume is compact,
the electric flux conservation requires
the total electric charge vanish.
If the integral of flux $F$ over the D3-brane boundary
is $2\pi N$, we need $N$ strings ending on the D3-brane worldvolume
to compensate the boundary charge.
This is also the case for a D3-brane ending on a $(k,1)$ fivebrane.

\begin{figure}[htb]
\centerline{\includegraphics{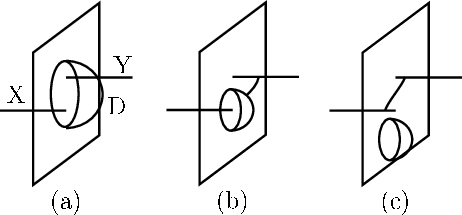}}
\caption{Quark-baryon transition}
\label{nqtob.eps}
\end{figure}

Baring this fact in mind,
we can show that
$N$ open strings
and a D3-brane disk can be continuously deformed
to each other.
In the following we treat three sets of D3-branes, and for distinction
we name them as follows:
\begin{itemize}
\item $X$ -- the coincident $N$ D3-branes between fivebranes $I$ and $I-1$.
\item $Y$ -- the coincident $N$ D3-branes between fivebranes $I$ and $I+1$.
\item $D$ -- a D3-brane disk whose boundary is ${\bf S}^2$ on fivebrane $I$.
\end{itemize}
We here assume that $N_I=N_{I-1}=N$.
Let us start from a D3-brane disk $D$ whose boundary is ${\bf S}^2$
on the fivebrane $I$ enclosing the both boundaries of $X$ and $Y$.
((a) in Figure \ref{nqtob.eps})
Although these boundaries carry magnetic charges coupled by $A$,
their charges cancel each other, and the net flux passing through
the boundary $\partial D$ is zero.
There are no open strings ending on $D$.

We move the disk so that $\partial Y$,
the boundary of $Y$,
gets out of $\partial D$.
When $\partial Y$ passes through $\partial D$,
the flux through $\partial D$ jumps by $N$,
and $N$ open strings stretched between $Y$ and $D$
are generated so that the total electric charge on the disk cancels.
((b) in Figure \ref{nqtob.eps})

If we keep moving the disk and $\partial X$ also gets out
of the boundary $\partial D$, the flux through the boundary jumps again by $-N$,
and this time $N$ open strings stretched between $D$ and $X$
are generated.
Two sets of $N$ strings can be connected to get off from $D$,
and we obtain $N$ open strings connecting $X$ and $Y$.
((c) in Figure \ref{nqtob.eps})
The disk can annihilate without any obstructions.

If $m_I=N_I-N_{I-1}\neq0$,
the D3-brane disk D in Figure \ref{nqtob.eps} (a)
is accompanied by $m_I$ strings attached on it.
This corresponds to the fact that we cannot define such 
$SU(N_{I-1}) \times SU(N_I)$ invariant operators as (\ref{baryonic1}) and (\ref{baryonic2})
due to the mismatch of the number of indices.
The $m_I$ open strings attached on the D3-brane disk corresponds to
$m_I$ fundamental or $-m_I$ anti-fundamental indices
which are not contracted.

\section{Three-form torsion and fractional branes}\label{torsion.sec}
In this section, we relate the fractional brane charge
and integrals of the $3$-form field on $3$-cycles.
Let us consider a process in which the number of the fractional branes
changes.
The fractional brane charge $Q\in H_3$
affects the $3$-form field $C_3$ and measured by
the integrals over $3$ cycles $\zeta$
\begin{equation}
\oint_\zeta C_3,\quad
\zeta\in H_3.
\end{equation}
We define the period integral at $r=r_0$ between the horizon $r=0$
and the AdS boundary $r=\infty$.
To change the fractional brane charge by $\Delta Q$, we add an M5-brane
wrapped on a $3$-cycle $\Delta Q\in H_3$ at
the AdS boundary, and move it to the horizon.
When the M5-brane pass through $r=r_0$, the period integrals
changes by
\begin{equation}
\Delta\oint_\zeta C_3=2\pi\langle\zeta,\Delta Q\rangle,
\label{deltac}
\end{equation}
where $\langle *,*\rangle$ is a map $H_3\times H_3\rightarrow U(1)$,
so called the torsion linking form, or, simply, the linking number.

The linking number is defined as follows.
Let $s$ be the order of $\zeta$.
Namely, $s$ is the smallest positive integer such that
$s\zeta$ is homologically trivial.
Such an integer always exists because $H_3$ is pure torsion.
There exists a $4$-chain $D$ such that
\begin{equation}
s\zeta=\partial D.
\end{equation}
We define the linking number $\langle\zeta,\eta\rangle$
of two $3$-cycles $\zeta$ and $\eta$ by
\begin{equation}
\langle\zeta,\eta\rangle=\frac{1}{s}\langle\langle D,\eta\rangle\rangle.
\end{equation}
where $\langle\langle D,\eta\rangle\rangle$ is the intersection number
of the $4$-chain $C$ and $3$-cycle $\eta$.
Because this number jumps by integers by continuous deformations,
only the fractional part of the linking number
is a topological invariant.

If we move an M5-brane wrapped on the $3$-cycle $\Delta Q$
from the AdS boundary to the horizon,
when it passes through $r=r_0$,
the M5-brane intersect with the $4$-chain $D$ at
$\langle\langle D,\eta\rangle\rangle$ points.
In this process, the four-form flux $G_4$ passing through
$D$, including the contribution of Dirac's string-like objects,
changes by
$2\pi\langle\langle D,\zeta\rangle\rangle$.
By using Stokes' theorem we obtain the relation
(\ref{deltac}).

For the manifold $M_{p,q,k}$, following the definition of the linking number,
we can easily obtain
\begin{equation}
k\langle {\bf v}_a,{\bf j}\rangle=m_a,\quad
k\langle {\bf w}_{\dot a},{\bf j}\rangle=m_{\dot a},
\label{linking0}
\end{equation}
for a general $3$-cycle ${\bf j}$ in (\ref{junc}).
The linking numbers among the basis are
\begin{equation}
\langle{\bf x}_a,{\bf x}_b\rangle=-\frac{1}{kq}\delta_{ab},\quad
\langle{\bf y}_{\dot a},{\bf y}_{\dot b}\rangle=-\frac{1}{kp}\delta_{\dot a\dot b},\quad
\langle{\bf x}_a,{\bf y}_{\dot b}\rangle=-\frac{1}{2kpq}.
\label{linking}
\end{equation}
Due to the constraint (\ref{summk}), the linking number among the basis
is not unique.
For example, a constant shift of all  the linking numbers in (\ref{linking})
does not affect the linking numbers for $3$-cycles which are
linear combination of the basis with the coefficient constrained by (\ref{summk}).

By ``integrating'' the relation (\ref{deltac})
and using (\ref{linking0}), we obtain
\begin{equation}
m_a-m_a^0=\frac{k}{2\pi}\oint_{{\bf v}_a} C_3,\quad
m_{\dot a}-m_{\dot a}^0=\frac{k}{2\pi}\oint_{{\bf w}_{\dot a}} C_3,
\label{deltacx}
\end{equation}
where $m_a^0$ and $m_{\dot a}^0$ are integration constants
which cannot be determined
from (\ref{deltac}).

Although gauge transformations can change the period integrals of $C_3$,
the relation (\ref{deltacx}) determines a element of $\Gamma/kH$ in a
gauge invariant way if we know $m_a^0$ and $m_{\dot a}^0$ because large gauge transformation change
the charge vector by an element of $kH$.

An important fact is that the constants $m_a$ and $m_{\dot a}$ depend on
the frame, the order of fivebranes.
The right hand side of the relations (\ref{deltacx}) are
defined on M-theory side, and is independent of the frame,
while $m_a$ and $m_{\dot a}$ on the left hand side
change by multiples of $k$ when
we change the order of fivebranes.
This means that $m_a^0$ and $m_{\dot a}^0$ depends on the frame,
and we cannot simply set them to be zero.

To obtain some information about the constants,
we use branes corresponding to baryonic operators.
Remember that in the IIB setup
baryonic operators correspond to D3-brane disks ending on
fivebranes,
and when $m_I\neq0$, they are accompanied by
$m_I$ open strings.

A similar phenomenon occurs on the M-theory side.
If there is non-trivial background $C$-field
M5-branes wrapped on five-cycles are accompanied by
M2-branes attached on their worldvolume,
and by identifying these M2-branes to strings
in the IIB setup,
we obtain relations between $m_I$ and
background $C$-field.

Let us consider the flux conservation
on M5-branes and how it relates the background $C$-field
and M2-branes attached on it.
The two-form field $b_2$ on M5-branes couples to the field strength $G_4$
in the bulk
by the coupling
\begin{equation}
S=\frac{1}{2\pi}\int_{M5}b_2\wedge G_4.
\end{equation}
This implies that the flux behaves as charge on M5-branes.
On the worldvolume of an M5-brane wrapped on a five-cycle
the total charge coupled by $b_2$ must cancel
due to the flux conservation.
This implies that, the cohomology class of the total charge
\begin{equation}
\left[\frac{1}{2\pi}G_4-\delta(\partial M2)\right]\in H^4(\Omega_I,{\bf Z})
\end{equation}
must be trivial.
$\delta(\partial M2)$ is the four-form delta function
with support on the boundaries of M2-branes.
By the Poincare duality, this is equivalent to
\begin{equation}
[g]=[\partial M2]\in H_1(\Omega_I,{\bf Z}),
\label{h1rel}
\end{equation}
where $g$ is the one-cycle Poincare dual to the flux $(2\pi)^{-1}G_4$.
The homologies $H_i(\Omega_a,{\bf Z})$ in the five-cycle are given by
\begin{equation}
H_0={\bf Z},\quad
H_1={\bf Z}_k,\quad
H_2={\bf Z}^{q-1},\quad
H_3={\bf Z}^{q-1}\oplus{\bf Z}_k,\quad
H_4=0,\quad
H_5={\bf Z}.
\label{homega}
\end{equation}
The homologies in $\Omega_{\dot a}$ are obtained by replacing $q$ in
(\ref{homega}) by $p$.
Because $H_1(\Omega_I,{\bf Z})={\bf Z}_k$ is pure torsion
we can rewrite (\ref{h1rel})
in terms of the linking form $H_3\times H_1\rightarrow U(1)$
as
\begin{equation}
\frac{1}{2\pi}\oint_\zeta C_3=\langle\zeta,\partial M2\rangle,
\label{m2bc}
\end{equation}
where $\zeta$ is the generator of the torsion subgroup of $H_3(\Omega_I,{\bf Z})$.
It is $\zeta={\bf v}_a$ for $\Omega_a$
and $\zeta={\bf w}_{\dot a}$ for $\Omega_{\dot a}$.
If we identify $m_I$ strings ending on a D3-brane disk
with a M2-brane wrapped on $m_I\gamma$
where  $\gamma$ is the generator of $H_1(M_{p,q,k},{\bf Z})=H_1(\Omega_I,{\bf Z})$,
(\ref{m2bc}) can be rewritten as
\begin{equation}
m_a=\frac{k}{2\pi}\oint_{{\bf v}_a} C_3,\quad
m_{\dot a}=\frac{k}{2\pi}\oint_{{\bf w}_{\dot a}} C_3\quad
\mod k.
\label{m2bc2}
\end{equation}
This means that
\begin{equation}
m_a^0=m_{\dot a}^0=0\mod k.
\end{equation}
This fixes only the frame independent part
of $m_a^0$ and $m_{\dot a}^0$.
Although in the $p=q=1$ case this reproduces the
result in \citen{Aharony:2008gk} for ABJM model,
this is not sufficient to establish
the relation between the fractional brane charge
and the $3$-form torsion for $p+q\geq3$.
We leave this problem for future works.

\section{Wrapped M2-branes and monopole operators}\label{wrappedm2.sec}
The correspondence between
Kaluza-Klein modes of massless fields in the internal manifold
and primary operators in the corresponding boundary CFT is
one of most important claim of AdS/CFT correspondence.

Such a correspondence
for ABJM model is discussed in \citen{Aharony:2008ug,Klebanov:2008vq}.
For more general ${\cal N}=2$ quiver gauge theories,
which describe M2-branes in toric Calabi-Yau $4$-folds,
the relation between
the holomorphic monomial functions, which are specified by
the charges of toric $U(1)$ symmetries,
and mesonic operators consisting of bi-fundamental fields,
was proposed in \citen{Lee:2007kv}.
In the reference, a simple prescription to
establish concrete coresspondence between Kaluza-Klein modes
and mesonic operators is given by utilizing
brane crystals\cite{Lee:2006hw,Lee:2007kv,Kim:2007ic}.
When this method was proposed,
it had not yet been realized that
the quiver gauge theories are actually quiver Chern-Simons theories.
After the importance of the existence of Chern-Simons terms
was realized,
this proposal was confirmed
\cite{Ueda:2008hx,Imamura:2008qs,Hanany:2008cd}
for special kind of brane crystals
which can be regarded as ``M-theory lift''
of brane tilings\cite{Hanany:2005ve,Franco:2005rj,Franco:2005sm}.

In three-dimensional spacetime, local operators in general carry
magnetic charges.
Such operators are called monopole operators.
In the correspondence between primary operators in three-dimensional CFT
and Kaluza-Klein modes,
monopole operators play an important role.
The results in \cite{Hanany:2005ve,Franco:2005rj,Franco:2005sm}
indicate
that the set of primary operators corresponding to the
supergravity Kaluza-Klein modes includes only a special kind of monopole
operators, ``diagonal'' monopole operators.
Diagonal monopole operators carries only the
diagonal $U(1)$ magnetic charges, and are constructed by combining
dual photon fields and chiral matter fields.
The concrete examples of diagonal operators have already apeared in
(\ref{bwtb}).
Because the canonical conjugate of the dual photon field
is the diagonal $U(1)$ field strength $F_D$,
the operator $e^{im\wt a}$ shifts the flux $F_D$ by $m$.

For a while we consider a generic Abelian
quiver ${\cal N}=2$ Chern-Simons
theory.
We label verices by $a$ and denote the corresponding gauge group by $U(1)_a$.
Let us consider a monopole operator with magnetic charges $m_a\in{\bf Z}$.
The diagonal monopole operator $e^{im\wt a}$
carries the same magnetic charge $m_a=m$ for all the
$U(1)_a$ gauge groups.
The gauge invariance of the operator requires
the Gauss law constraint
\begin{equation}
m_ak_a+Q_a=0,
\end{equation}
where $Q_a$ is the $U(1)_a$ electric charge
carried by matter fields included in the monopole operator.
This guarantees the invariance of the operator under the gauge symmetry
(\ref{agauge}).
By summing up this over all $a$,
we obtain the constraint
\begin{equation}
\sum_am_ak_a=0.
\end{equation}
Therefore,
monopole operators are labeled by $n-1$ independent magnetic charges.
One of them is the diagonal monopole charge,
and corresponds to a certain component of Kaluza-Klein momentum in the internal space
(the D-particle charge from the type IIA perspective).

What are the interpretation of the other $n-2$ magnetic charges?
It is natural to identify these with the
charges of M2-branes wrapped on two-cycles.
Let us return to the ${\cal N}=4$ Chern-Simons theory studied in this paper,
which is a special case of ${\cal N}=2$ quiver Chern-Simons theories.
The two-cycle homology of the corresponding internal space $M_{p,q,k}$
is
\begin{equation}
H_2(M_{p,q,k},{\bf Z})={\bf Z}^{p+q-2},
\end{equation}
and the Betti number coincides with the number of independent
magnetic charges of non-diagonal monopole operatords.

We now explain why we did not impose $G_B$ gauge invariance
on baryonic operators.
First, let us remember 
the reason why symmetry groups which act on
wrapped branes are usually regarded as global symmetries.
Consider $AdS_{d+1}$
with the metric
\begin{equation}
ds^2=\frac{R^2}{z^2}((dx^\mu)^2+dz^2),\quad
\mu=1,\ldots,d,
\end{equation}
and let $A_\nu(x^\mu,z)$ be a $U(1)$ gauge field
coupling to wrapped branes.
We follow \citen{Klebanov:1999tb} and consider the Euclidian AdS space.
$z$ is the radial coordinate such that the AdS boundary is at $z=0$.
Let us assume the asymptotic behavior of the vector field as
\begin{equation}
A_\nu\propto z^\Delta.
\label{aiasym}
\end{equation}
For the convergence of the Euclidian action,
$\Delta$ must satisfy the inequality
\begin{equation}
\frac{d}{2}-2<\Delta.
\label{ineq}
\end{equation}
With the equation of motion $d*dA=0$ we obtain
the asymptotic behavior of the gauge field
\begin{equation}
A_\nu(x^\mu,z)=a_\nu(x^\mu)+z^{d-2}b_\nu(x^\mu).
\label{aiasym2}
\end{equation}
On the AdS boundary we need to impose boundary condition which
fixes one of $a_\nu(x^\mu)$ and $b_\nu(x^\mu)$.
When $d\geq4$, only the second term in (\ref{aiasym2}) is
allowed by (\ref{ineq})
and the boundary condition $a_\nu(x^\mu)=0$ must be
imposed.
Then the gauge field asymptotically vanishes near the boundary,
and this is the reason why the symmetry is global in the boundary CFT.

On the other hand, when $d=3$, both terms in
(\ref{aiasym2}) satisfy the inequality
(\ref{ineq}), and we can choose any one of
$a_\nu(x^\mu)=0$ (Dirichlet)
and $b_\nu(x^\mu)=0$ (Neumann)
as the boundary condition.
Indeed, these two boundary conditions first appeared in \citen{Breitenlohner:1982jf} 
and are used in \citen{Witten:2003ya} to construct a pair of Chern-Simons theories
which are ``S-dual'' to each other.
Let us take the Neumann boundary condition.
In this case, 
the boundary value
of the gauge field $a_\nu(x^\mu)=A_\nu(x^\mu,z=0)$ does not vanish, and
is dynamical in the sense that it is path integrated.
Thus we can regard this as a gauge field in the boundary CFT,
and the wrapped branes coupled by $A_\nu$ should be cherged objects
in the boundary CFT, too.
Because the Dirichlet and Neumann boundary conditions
are exchanged by the duality transformation of the gauge field, 
both kinds of operators corresponding to electric and
magnetic particles in the AdS$_4$ cannot be gauge invariant .

In the case of our M-theory background,
the gauge fields $A^i=A_\nu^idx^\nu$ coupling to wrapped M5-branes
and $\wt A_i=A_{i\nu}dx^\nu$ coupling to wrapped M2-branes
are defined by
\begin{equation}
C_6=\sum_{i=1}^{p+q-2}\omega_i\wedge A^i,\quad
C_3=\sum_{i=1}^{p+q-2}\omega^i\wedge\wt A_i,
\end{equation}
where $C_3$ and $C_6$ are the three- and six-form potential field,
which are dual to each other,
and $\omega_i$ and $\omega^i$ are cohomology basis of
$H^5(M_{p,q,k},{\bf Z})={\bf Z}^{p+q-2}$
and
$H_{\rm free}^2(M_{p,q,k},{\bf Z})={\bf Z}^{p+q-2}$,
respectively. 
Because $A^i$ and $\wt A_i$ are electric-magnetic dual to each other,
it is impossible to impose the Dirichlet boundary condition on all
of them, and consequently some of wrapped branes inevitablly correspond to
gauge variant operators.
This is the reason why we did not require the baryonic operators
to be $G_B$ gauge invariant.
Although it may be possible to take some S-dual picture in which
wrapped M5-branes correspond to gauge invariant operators,
then we have to relate wrapped M2-branes to gauge variant operators.

\section{Conclusions}
In this paper, we investigated some aspects in the gravity dual of
${\cal N}=4$ quiver Chern-Simons theories.
One is fractional branes.
We confirmed that the group of fractional brane charge,
which is obtained by the analysis of Hanany-Witten effect in
the type IIB brane configuration,
is isomorphic to the homology $H_3(M_{p,q,k},{\bf Z})$.
We also established the relation between the fractional brane charge
and the torsion of the $3$-form field
up to the frame dependent constants.
In order to determine the constant part,
more detailed analysis would be needed.

We also discuss
the duality between baryonic operators in the Chern-Simons theory
and M5-branes wrapped on five-cycles in $M_{p,q,k}$.
We defined baryonic operators which carries $G_B$ charges,
and found that the homology group $H_5(M_{p,q,k},{\bf Z})$
is consistent with
the decomposability of products of baryonic operators into mesonic ones
on the field theory side.
We also found that the conformal dimension of baryonic operators are
consistent with the mass of the wrapped M5-branes.
The degeneracy of the baryonic operators were explained
as the degeneracy of the ground states for the
collective motion of the wrapped M5-branes.

We also commented on the relation between non-diagonal
monopole operators and wrapped M2-branes.
The two-cycle Betti number $b_2$ of the
internal manifold $M_{p,q,k}$ is found to coincides with
the number of independent magnetic charges of non-diagonal
monopole operators.

We did not impose the gauge invariance on baryonic operators.
In Section \ref{wrappedm2.sec} we showed that
some of wrapped M2-branes and wrapped M5-branes
inevitablly correspond to gauge variant operators
in the boundary CFT.

There are many questions left which should be studied.
The extension of our analysis to
more general quiver Chern-Simons theories with smaller
supersymmetry
is one of them.
Moduli spaces of ${\cal N}=2$ supersymmetric quiver Chern-Simons theories
are studied in
 \citen{Martelli:2008si,Ueda:2008hx,Imamura:2008qs,Hanany:2008cd}.
For the class of theories which described by brane tilings \cite{Hanany:2005ve,Franco:2005rj,Franco:2005sm}
(See also \citen{Kennaway:2007tq,Yamazaki:2008bt} for reviews.)
there is a simple prescription to establish the relation between
toric data of Calabi-Yau $4$-folds
and Chern-Simons gauge
theories \cite{Ueda:2008hx,Imamura:2008qs,Hanany:2008cd}.
It may be interesting to extend our analysis to such a large class of
theories.

In general, dual CFT of toric Calabi-Yau $4$-folds
cannot be described by brane tilings.
In such a case, brane crystals \cite{Lee:2006hw,Lee:2007kv,Kim:2007ic}
are expected to play an important role.
The relation between brane crystals and dual CFT
are not fully understood, and the analysis of
homologies and wrapped branes
may be helpful to obtain some information about dual CFT.

We hope we will return to these subjects in near future.

\section*{Acknowledgements}
We would like to thank K.~Kimura, T.~Watari 
and F.~Yagi for valuable discussions.
We also thank Y.~Tachikawa for a helpful comment.
Y.~I. is partially supported by
Grant-in-Aid for Young Scientists (B) (\#19740122) from the Japan
Ministry of Education, Culture, Sports,
Science and Technology. S.Y. is supported by 
Global COE Program "the Physical Sciences Frontier", MEXT, Japan.

\end{document}